\documentstyle[12pt,epsf]{article}
\begin{document}

\begin{center} {\bf  Directed polymers on a Cayley tree with spatially
correlated disorder}

\vspace {.2in}

by\\

\vspace{.2in}

Yadin Y. Goldschmidt\\
Department of Physics and Astronomy\\
University of Pittsburgh\\
Pittsburgh, PA  15260

\end{center}

\vspace{.3in}

\begin{center}
  {\bf {Abstract}}
\end{center}

In this paper we consider directed walks on a tree with a fixed
branching ratio K at a finite temperature T.  We consider the case
where each site (or link) is assigned a random energy uncorrelated in
time, but correlated in the transverse direction i.e. within the
shell. In this paper we take the transverse distance to be the
hierarchical ultrametric distance, but other possibilities are
discussed. We compute the free energy for the case of quenched
disorder and show that there is a fundamental difference between the
case of short range spatial correlations of the disorder which behaves
similarly to the non-correlated case considered previously by Derrida
and Spohn and the case of long range correlations which has a totally
different overlap distribution which approaches a single delta function
about $q=1$ for large $L$, where $L$ is the length of the walk. In the
latter case the free energy is not extensive in $L$ for the
intermediate and also relevant range of $L$ values, although in the true
thermodynamic limit extensivity is restored. We identify a
crossover temperature which grows with $L$, and whenever $T<T_c(L)$
the system is always in the low temperature phase. Thus in the case of
long-ranged correlation as opposed to the short-ranged case 
a phase transition is absent.

\newpage
\section{{\bf Introduction}}
\label{sec:1}

The problem of directed polymers in a random medium can be formulated
on a lattice \cite{huse,kardar1,c&d1,mezard,halpin-zhang}, or in the 
continuum limit \cite{kardar2,parisi1,mp,fisher-huse,BYG,YYG,YGB}.  On
a lattice there is a random energy associated with each bond (or
site). Walks (or polymers) start at a given point and are allowed to
proceed only along the positive direction of one of the coordinates
which is referred to as ``time''. The other coordinates are referred to
as ``transverse''. The partition function is given by
\begin{eqnarray}
Z_L(\beta)\ =\ \sum_w e^{-\beta E(w)}\ ,
\label{PF}
\end{eqnarray}
where the sum is over all walks $w$ of $L$ steps and 
\begin{eqnarray}
E(w)\ =\ \sum_{(ij)\in w} \epsilon_{ij}
\label{E}
\end{eqnarray}
is the sum of the random energies along the walk. $\beta = 1/T$ is the
the inverse temperature in the proper units.

In the continuum limit the partition function is given by the
functional integral 
\begin{eqnarray}
Z(\beta)\ =\ \int [D{\bf x}(t)] \exp \left\{ -\beta \int_0^L dt 
\ \left[ {1 \over 2}
\dot{{\bf x}}(t)^2 \ + \ V({\bf x}(t),t) \right] \right\} ,
\label{ZC}
\end{eqnarray}
where ${\bf x}(t)$ is the ($d$-1)-dimensional transverse position of the
polymer at time $t$ and $V({\bf x}(t),t)$ is the random potential. The
term $(1/2) \dot{{\bf x}}(t)^2$ measures the bending energy of the
polymer.

On the special lattice of a Cayley tree and with uncorrelated disorder
\begin{eqnarray}
\langle \epsilon_{ij} \epsilon_{lm}  \rangle \ = \ g \delta_{il} \delta_{jm},
\label{0c}
\end{eqnarray}
many properties of the model could be extracted analytically \cite{DS,BPP},
like the free energy and the probability of overlaps between two walks. 
The model exhibits a phase transition at finite temperature $T_c$. 
Define $q(w,w')$ to be the fraction of their length that two walks
$w,\ w'$ of length $L$ spend together. The probability distribution
for overlaps is then given by
\begin{eqnarray}
P(q)=\left\langle {1 \over Z_L^2}\sum_w \sum_{w'} \delta(q-q(w,w'))\ 
\exp(- \beta E_w - \beta E_{w'})\right\rangle.
\label{Pqdef}
\end{eqnarray}
For $T>T_c$ it was found that the probability of overlaps is a single
delta function at $q=0$, whereas for $T<T_c$ the
probability distribution consists of a weighted sum of two delta
functions:
\begin{eqnarray}
P(q)\ = \ {T \over T_c} \delta(q)\ + \ \left(\ 1\ -\ {T \over T_c}\ 
\right)\ \delta(q-1).
\label{Pq}
\end{eqnarray}
This distribution implies that the free energy landscape consists of
many valleys separated by large barriers. Two walks lying in the same
valley have an overlap of $q=1$, whereas walks lying in different
valleys have zero overlap.
 
In the continuum limit the model has been treated for general $d$ by
the variational approximation \cite{mp} and by the $1/d$-expansion
\cite{YYG}, both  valid for large $d$. For the special case of $d=2$ a
Bethe ansatz technique yielded some exact results \cite{kardar2}. In the
continuum limit both the case of short-ranged and of long-ranged
correlations of the disorder have been considered. The correlations
have been defined by 
\begin{eqnarray} 
\langle V({\bf x},t)\ V({\bf x}',t') \rangle\ = \ \delta(t-t')\ 
F(({\bf x}-{\bf x}')^2), 
\label{VV} 
\end{eqnarray} 
and classified as short-ranged or long-ranged according to the form of
$F(y^2)$. For the short-ranged case one usually takes \cite{mp,engel,horner}  
\begin{eqnarray} 
F(y^2) = {g \over {\gamma-1}} \ ( a_0+y^2)^{(1-\gamma)},
\label{Fys} 
\end{eqnarray} 
where $g>0$ is the strength of the disorder and
$\gamma>2$ determines the range of the correlations.  

The case of $1<\gamma <2$ is considered long-ranged. Also considered 
long-ranged are correlations of the form 
\begin{eqnarray}
F(y^2) = a_0\ - \ {g \over {1-\gamma}} \ y^{2(1-\gamma)},
\label{Fyl}
\end{eqnarray}
with $0 \leq \gamma<1$. The constant $a_0$ is important to maintain
the requirement $|F(y^2)| \leq F(0)$ dictated by the Schwarz inequality
for the appropriate $y$-range,
but is sometimes neglected in the literature \cite{mp}, likely because it
contributes only a trivial constant to the free energy. 
In the case of a correlation of the form (\ref{Fyl}), $a_0$ has to be
very big, such that $[a_0(1-\gamma)/g]^{1/(2(1-\gamma))}$ is greater
than the system size. An example of this kind of correlations is the
so-called random field case (see e.g. \cite{engel}), for which $\gamma=1/2$ and
\begin{eqnarray}
F(y^2)=h^2 (N- |y|),
\end{eqnarray}
where $N$ is the system size. 
In the short-ranged case a phase transition in terms of the
temperature (or the strength of the disorder) has been found, where
the two phases differ in the nature of the distribution of overlaps. A
one-step replica-symmetry-breaking solution has been found for $T<T_c$
when using the variational approximation. Both phases in this
approximation were found to be characterized by the trivial wandering
exponent $\nu=1/2$ defined by
\begin{eqnarray}
\left\langle \ \overline{{\bf x}^2(L)} \right\rangle \propto L^{2\nu}
\label{nu}
\end{eqnarray}
(here the bar represents configurational or thermal average and the
brackets refer to averaging over the disorder). This is unlike the
exact analytical result at $d=2$ which yields $\nu=2/3$
(superdiffusion) \cite{huse,kardar2}, and simulations that found
$\nu>1/2$ for $d \geq 2$ \cite{kim,nissila}. There were claims in the
literature that the wandering exponent is greater than $1/2$ for any
finite $d$ in the disordered dominated phase \cite{nissila}. On the
other hand there are claims that have gained more momentum recently
\cite{c&d2,YYG,frey,moore,lassig}, that the exponent becomes trivial
(1/2) at an upper critical dimension, which is presumably $d_c=5$
(four transverse directions).

For the case of long-ranged correlations the variational approximation
for the continuum limit model predicts \cite{mp}
\begin{eqnarray}
\nu_{ _{Flory}}\ = \ { 3 \over 2(1+\gamma)}\ \ ,
\label{Flory}
\end{eqnarray}
and it is not known if this result is exact at any finite
dimension. In this case no phase transition is 
found as a function of the temperature (or the strength of the
disorder) and there is an infinite-step replica-symmetry-breaking
(RSB) solution ($\grave{a}$ la Parisi) for the appropriate order
parameter, which manifests itself as a continuous (non-delta-function)
part in the overlap distributions for the walks.

The special case of long-ranged harmonic correlations ($\gamma=0$) has
been solved exactly by Parisi \cite{parisi2}. An inspection of his solution
reveals that in that case the free energy is not extensive but grows
as $L^2$ (where $L$ is the size of the system in the temporal
direction). Another exponent which is often referred to in the
literature is the exponent characterizing the free energy fluctuations
\begin{eqnarray}
\langle F^2 \rangle \ -\ \langle F \rangle^2 \propto L^{2 \omega}.
\label{omega}
\end{eqnarray}
The $\omega$ exponent is related to $\nu$ through the scaling relation
\begin{eqnarray}
2\ \nu\ = \ 1\ +\ \omega.
\label{scaling}
\end{eqnarray}
Thus for the harmonic case $\nu=3/2$ and $\omega=2$. The harmonic case
is special in the sense that its solution is replica symmetric. 

The case of long-ranged correlations, or in fact any non-zero
correlations of the disorder, has not been investigated in the
previous treatments of the model on the Cayley tree. The interesting
results obtained for the long-ranged case in the continuum limit, and
the fact that an exact solution has been found in the case of harmonic
correlations, motivated us to carry out an investigation of the case
of non-zero ranged spatial correlations on a Cayley tree, both for the
case of short and long-ranged correlations.  We obtain some
interesting results which will be presented below together with some
open questions. A thorough understanding of the directed polymer
problem is particularly important because of its connection with the
KPZ equation \cite{KPZ} and with the behavior of flux lines in
high-$T_c$ superconductors \cite{blatter,ledoussal,ertas,goldschmidt}.
There is also a well-known mapping from the directed polymer problem
to Burgers' turbulence, where the case of long-ranged correlations is
of importance \cite{BMP}.

\section{{\bf The tree problem}}
\label{sec:2}
We consider directed walks on a branch of a Cayley tree of
coordination number $K+1$ (see Fig. 1). Each bond branches into K new
bonds in the forward ``time'' direction. For each site (or
alternatively each bond ending at the given site) we choose a random
energy $\epsilon(t,z)$ where $t$ designates the shell, $0\leq t \leq
L$, and $z$ is a label within the shell that can take $K^t$ values. The
random energies are chosen from a gaussian distribution satisfying
\begin{eqnarray}
\langle \epsilon(t,z) \rangle\ &=& \ 0  \label{e} \\
\langle \epsilon(t,z)\ \epsilon(t',z')  \rangle \ &=&\ \delta_{t,t'}\
f(d(z,z')). 
\label{ee}
\end{eqnarray}
Here $f(d)$ is a function to be specified later, and $d(z,z')$ is a
distance among points (sites) belonging to the same shell. Energies at
different shells are uncorrelated.

In this paper we choose the following definition for $d(z,z')$:
\begin{eqnarray}
d_u(z,z')\ =&&{\rm number\ of\ steps\ for\ two\ walks\ starting\ at\
  z\ and\ z'} \nonumber \\ 
&&{\rm and\ moving\ \underline{backwards}\ in\ time\ to\ meet}.
\label{du}
\end{eqnarray}
This is a hierarchical distance between two points. It also satisfies
$d_u(z,z)=0$ and $0\leq d_u(z,z')\leq t$ within the $t$-shell. This
distance is also referred to as an ultrametric distance \cite{MPV} (hence the
subscript $u$), since it satisfies a stronger inequality than the
ordinary triangular inequality, namely
\begin{eqnarray}
d_u(z,z')\ \leq\ \max (d_u(z,y),\ d_u(y,z')),
\label{ultara}
\end{eqnarray}
for any point $y$ within the shell.
This ultrametric distance is very different from an euclidean distance
on lattices characterized by translational invariance in real space,
but is sufficient for defining spatial correlations of the disorder
within a shell, and make the problem amenable to an exact solution.

A different choice for a distance which is more suitable for
calculating the root-mean-square transverse distance can be defined as
follows \cite{c&d2}. Let us label each branch of the tree by $1, \cdots,
K$, which we call directions. These labels are a priori arbitrary on a
tree, but once the choice is made it remains fixed at each branching
point. For a given walk of length $t$ starting at the origin we denote
by $z_1$ the number of times the walk moves in the $1$-direction, by
$z_2$, the number of times it moves in the $2$-direction, etc. We then
associate a vector $(z_1-t/K,\cdots ,z_K-t/K)$ with the end point of
the walk on the $t$-shell. We denote this $K$-dimensional vector by
{\bf R}(z). Note that there is not a one-to-one correspondence between
points $z$ on the tree and vectors ${\bf R}$ as there are different
points which are associated with the same vector. 
The transverse distance between two points is defined as
\begin{eqnarray}
d_{tr}(z,z')\ =\ \left( \sum_{i=1}^K (R_i(z)-R'_i(z'))^2 \right)^{1/2}.
\label{dtr}
\end{eqnarray}
This distance always satisfies the inequality
\begin{eqnarray}
d_{tr}(z,z') \leq \sqrt{2} d_u(z,z'),
\label{ineq}
\end{eqnarray}
for any two points $z,\ z'$. The advantage of this distance is that in
the absence of disorder one has
\begin{eqnarray}
\overline{{\bf R}}\ &=&\ 0 \\
\overline{{\bf R}^2}\ &=&\ {1 \over K}\left(1-{1 \over K}\right) L,
\label{R2}
\end{eqnarray}
where the bar denotes configurational average over all walks of length
$L$. Thus $\nu=1/2$ as is expected for a random walk. In the presence
of disorder with spatial correlations, this distance is harder to use
in a calculation of the quenched free energy of the model, and further
discussion of the use of this distance will be given in a future
publication.

\section{{\bf The replica solution}}
\label{sec:3}

The method we use in this section is a generalization of the method
used in Appendix 1 of ref. \cite{c&d2} for uncorrelated disorder. To
calculate the free energy of the model we use the replica trick
\begin{eqnarray}
-\beta F \ =\ \lim_{n \rightarrow 0} { 1 \over n } \ln \langle Z_L^n \rangle.
\label{fe}
\end{eqnarray}
We can think of $Z_L^n$ as the partition function of $n$ different
walks of length $L$ emanating from the origin of a branch of a tree.
Adopting Parisi's scheme for RSB in real space, we assume that the
following arrangement of the $n$ walks gives the leading contribution
to the free energy in the $n \rightarrow 0$ limit when $L$ is large:
\begin{itemize}
\item (a) The $n$-walks stay together for the first $L(q_1-q_0)$ steps
(where $q_0=0$ and $0 \leq q_1 \leq 1$).
\item (b) The walks split into $m_1$ bundles of $(n/m_1)$ walks each
and remain so for a time $L(q_2-q_1)$.
\item (c) Continuing in this way, in the j'th step the walks split
into $m_j$ groups each comprising of $(n/m_j)$ walks and remain so for
a time $L(q_{j+1}-q_j) \equiv L \Delta q_j$.
\item (d) Finally, at time $t_M=Lq_M=L$, the walks split into $n$
individual walks.
\end{itemize}
Thus
\begin{eqnarray}
0\ &=&\ q_0\leq q_1\leq \cdots \leq q_M \ = \ 1 \\
1\ &=&\ m_0\leq m_1\leq \cdots \leq m_M \ = \ n.
\label{qm}
\end{eqnarray}
We also define
\begin{eqnarray}
x(q_j)=\ n/m_j
\label{xq}
\end{eqnarray}
and thus
\begin{eqnarray}
1\ =\ x(q_M) \leq \cdots \leq x(q_j) \cdots \leq x(q_0)=n .
\label{xq1}
\end{eqnarray}
For large $L$, $\langle Z_L^n \rangle$ is given by
\begin{eqnarray}
\langle Z_L^n \rangle\ =\ \max_{\{q_j\}} \max_{\{m_j\}}
\prod_{j=0}^{M-1} \left( K^{m_j} \left\langle \exp\left\{-\beta (n/m_j)\ 
(\epsilon_1^{(t)}+\cdots \epsilon_{m_j}^{(t)})\right\}  \right\rangle 
\right)^{L\Delta q_j}.
\label{ZLn}
\end{eqnarray}
Here $\epsilon_i^{(t)}$ denotes the energy encountered by a walker
belonging to the $i$'th group at time $t$.

The factorization in eq. (\ref{ZLn}) follows from the fact that random
energies at different times are uncorrelated. The factor $K^{m_j}$ is
a geometrical degeneracy factor which is entropic in origin. It
arises because each of the $m_j$ groups can choose its next step among
$K$ different possibilities, and each possibility gives rise, as will
become clear in the sequel, to a configuration characterized by the
same contribution to the partition function (same Boltzmann weight).
One can also associate a combinatorial factor with different
possibilities to assign individual walks to bundles when they split,
but this turns out to give a total factor of $n!$, which becomes 1 in
the $n \rightarrow 0$ limit.  

To proceed we use the fact that the random energies are chosen from a
gaussian distribution satisfying equation (\ref{ee}). It then follows
that
\begin{eqnarray}
\left\langle \exp\left\{-\beta (n/m_j)\ 
(\epsilon_1^{(t)}+\cdots \epsilon_{m_j}^{(t)})\right\} \right\rangle
\ &=&\ \exp\left\{{1 \over 2}\beta^2 (n/m_j)^2 \ \left\langle
(\epsilon_1^{(t)}+\cdots \epsilon_{m_j}^{(t)})^2 \right\rangle
\right\}, \label{avboltz} \\
\left\langle (\epsilon_1^{(t)}+\cdots \epsilon_{m_j}^{(t)})^2
\right\rangle \ &=&\ m_j f(0)\ +\ 2\sum_\ell N_{j,\ell} f(\ell)
\label{ave2}
\end{eqnarray}
where $N_{j,\ell}$ is the number of pairs of distance $\ell$ at time
$t \in [t_j,t_{j+1}]$ among the $m_j$ groups of walkers, where
$t_j=Lq_j$. 
The coefficients $N_{j,\ell}$ satisfy
\begin{eqnarray}
m_j \ +\ 2\sum_\ell N_{j,\ell}\ =\ m_j^2.
\label{sumN}
\end{eqnarray}
A careful enumeration of the distribution of distances after $t$
steps ($t_j<t<t_{j+1}$) reveal that the following identity holds:
\begin{eqnarray}
N_{j,L(q_j-q_k)+\Delta t}\ =\ 
{1 \over 2} \left({1 \over m_{k-1}}-{1\over m_k}\right) m_j^2, \ \ \ \
k=1, \cdots, j
\label{Njell}
\end{eqnarray}
with $\Delta t=t-t_j$. Since we are interested in the limit $M
\rightarrow \infty$, i.e. $\Delta q \rightarrow 0$, we will take
$N_{j,\ell}$ to depend only on j and omit $\Delta t$ in
eq. (\ref{Njell}).  
In deriving eq. (\ref{Njell}) we used the fact that the groups of
walkers split at each time $t_j=Lq_j$ according to the procedure
described at the beginning of the section, as well as the definition of
the hierarchical (ultrametric) distance. 
Substituting the result (\ref{Njell}) in eq. (\ref{ave2}) we find
\begin{eqnarray}
(n/m_j)^2 \ \left\langle (\epsilon_1^{(t)}+\cdots
\epsilon_{m_j}^{(t)})^2 \right\rangle \ =\ n
x(q_j)f(0)\ +\ n \sum_{k=1}^j (x(q_{k-1})-x(q_k))
f(L(q_j-q_k)).
\label{ave2e}
\end{eqnarray}
In the limit $n \rightarrow 0$ the inequalities (\ref{qm}) and
(\ref{xq1}) are inverted and hence
\begin{eqnarray}
0\ =\ x(q_0) \leq \cdots \leq x(q_j) \cdots \leq x(q_M)=1
\label{xq2}
\end{eqnarray}
Using expression (\ref{ave2e}) in eq. (\ref {avboltz}) and
subsequently in the 
formula for $\langle Z_L^n\rangle$, eq. (\ref{ZLn}), we obtain
\begin{eqnarray}
{1 \over nL} \ln \langle Z_L^n\rangle\ &=&\ \sum_{j=0}^{M-1} {\Delta
q_j \over x(q_j)} \ln K\ +\ \frac{1}{2}\beta^2\sum_{j=0}^{M-1}\Delta
q_j x(q_j) f(0) \nonumber \\ &-&\
\frac{1}{2}\beta^2\sum_{j=0}^{M-1}\Delta q_j \sum_{k=1}^j \Delta q_k
\left({x(q_k)-x(q_{k-1}) \over \Delta q_k}\right) f(L(q_j-q_k)),
\label{Fdiscr}
\end{eqnarray}
where an extremum over $x(q_j)$ has to be taken. In the limit of large
$M$, $q$ becomes a continuous variable in the interval [0,1], and
$x(q)$ becomes a function on that interval satisfying $0\leq x(q) \leq
1$. The summations in eq. (\ref{Fdiscr}) become integrals and we have
\begin{eqnarray}
{-\beta F \over L}\ =\ \int_0^1 {dq \over x(q)}\ \ln K\ +\
\frac{1}{2}\beta^2\int_0^1 dq\ x(q)\ f(0)\ -\
\frac{1}{2}\beta^2\int_0^1 dq\ \int_0^q dp\ x'(p)\ f(L(q-p)),
\label{bFint}
\end{eqnarray}
where $x'(p)$ stands for $dx/dp$. This expression can be further
simplified: First we perform the $p$ integration by parts 
using the fact that x(0)=0. Next, in the second term which 
remains a double integral we change the order of integration 
\begin{eqnarray}
\int_0^1 dq \int_0^q dp = \int_0^1 dp \int_p^1 dq,
\label{chord}
\end{eqnarray}
leading finally to a single integral. These steps yield
\begin{eqnarray}
{-\beta F \over L}\ =\ \int_0^1 {dq \over x(q)}\ \ln K\ +\
\frac{1}{2}\beta^2\int_0^1 dq\ x(q)\ [ f(0) - f(L(1-q)) ] ,
\label{bFfin}
\end{eqnarray}
where $x(q)$ is to be determined by extremizing this expression.
Equation (\ref{bFfin}) is the main result of this section.

Before we end this section let us mention a simple generalization of
the problem. Equation (\ref{ee}) can be replaced by a more general form
\begin{eqnarray}
\langle \epsilon(t,z)\ \epsilon(t',z') \rangle \ &=&\ \delta_{t,t'}\
f(t,d(z,z')).
\label{eem}
\end{eqnarray}
so the spatial correlation can vary with time (but energies at
different times are still uncorrelated). The possibility of a time
dependent width of the distribution of disorder in the uncorrelated
case has been considered in ref. \cite{DS}.
Using eq. (\ref{eem}), we can repeat
all the steps leading eq. (\ref{bFint}), and we find that it is replaced by
\begin{eqnarray}
{-\beta F \over L}\ =\ \int_0^1 {dq \over x(q)}\ \ln K\ +\
\frac{1}{2}\beta^2\int_0^1 dq\ x(q)\ f(Lq,0)\nonumber \\
 -\ \frac{1}{2}\beta^2\int_0^1 dq\ \int_0^q dp\ x'(p)\ f(Lq,L(q-p)).
\label{bFt}
\end{eqnarray}
It is easy to pursue this more general case by the same method
presented in the next section, but we will not consider it further in
this paper.

\section{ {\bf Solution for the free energy and overlap distribution }}
\label{sec:4}

In order to extremize the expression for the free energy (\ref{bFfin})
derived in the last section, we take a functional derivative with
respect to $x(q)$ to obtain
\begin{eqnarray}
-{\ln K \over x^2(q)}\ +\ \frac{1}{2}\beta^2(f(0)-f(L(1-q))\ = \ 0,
\label{}
\end{eqnarray}
and hence
\begin{eqnarray}
x(q)\ =\ {T\sqrt{2\ln K} \over \sqrt{f(0)-f(L(1-q))}}
\label{xqext}
\end{eqnarray}
This solution is valid for the range of $q$ for which the inequality
$0\leq x(q)\leq 1$ holds. Otherwise one has to choose $x(q)$ at its
maximal (or minimal) allowed values.

Before we consider some concrete candidates for the spatial
correlation function, we comment about the posssible admissible
correlations. The Schwarz inequality requires that the function
$f(y)$ defined in equation (\ref{ee}) satisfy $|f(y)| \leq f(0)$. More
generally for the k'th level the covariance matrix
\begin{eqnarray}
C_{z,z'}=\langle \epsilon(k,z)\epsilon(k,z')\rangle=f(d(z,z')),
\end{eqnarray}
must be positive semi-definite, which means that all its eigenvalues
are non-negative. In the k'th shell the hierarchical distance takes
values from 1 to k. Writing down the covariance matrix for the k'th
shell we verified that a sufficient condition for an admissible
correlation of the disorder is a function $f(y)$ which is always
non-negative and is a monotonically decreasing function of the
hierarchical distace $y$. For simplicity we will limit the discussion
for the case $K=2$, but it can be generalized to any value of $K$.
First we present as an example the eigenvalues of the covariance
matrix in the 3'd shell of the BL:
\begin{eqnarray*}
f(0)-f(1),  \ \ \ \rm{(multiplicity\ 4)} \\
f(0)+f(1)-2f(2), \ \ \ \rm{(multiplicity\ 2)} \\
f(0)+f(1)+2f(2)-4f(3), \\
f(0)+f(1)+2f(2)+4f(3),
\end{eqnarray*}
and these are all non-negative if $f(0)>f(1)>f(2)>f(3)\geq 0$. In
general we now obtain by induction all the eigenvalues of the
covariance matrix of the (k+1)'th shell once we know the eigenvalues
of the matrix of the k'th shell. 
For k=1 the covariant matrix is given by
\begin{eqnarray}
\left(\begin{array}{ll}f(0)\ & f(1) \\
f(1)\ & f(0) \end{array} \right)
\end{eqnarray}
which has eigenvectors (1,-1) and (1,1) and corresponding eigenvalues
$f(0)-f(1)$ and $f(0)+f(1)$. For k=2 the covariant matrix is a 4 by 4
matrix
\begin{eqnarray}
C(2)=\left(\begin{array}{ll}C(1)\ & D(1) \\ D(1)\ & C(1) \end{array}
\right),
\end{eqnarray}
where $C(1)$ is the k=1 covariance matrix and $D(1)$ is a 2 by 2
matrix, all the elements of which are equal to $f(2)$. The eigenvectors
of $C(2)$ are (1,-1,0,0) and (0,0,1,-1) with the corresponding
eigenvalue $f(0)-f(1)$ of multiplicity 2. In addition there are two new
eigenvectors (1,1,-1,-1) and (1,1,1,1) with eigenvalues
$f(0)+f(1)-2f(2)$ and $f(0)+f(1)+2f(2)$.
In the k'th level there are always
k+1 distinct eigenvalues, the (k+1)'th of which is always given by
\begin{eqnarray}
\lambda_{k+1}(k)=f(0)+f(1)+2f(2)+ \cdots +2^{k-1}f(k).
\end{eqnarray}
At level (k+1) the first k eigenvalues remain the same as level k
first k eigenvalues, but with double multiplicity. 
This is because the corresponding eigenvectors are just obtained from
the previous eigenvectors by adding zeros at the beginning or at the
end.
But there are two new eigenvectors $(1,\cdots,1,-1,\cdots,-1)$ and
$(1,\cdots,1)$ with corresponding eigenvalues which are given by
\begin{eqnarray}
\lambda_{k+1}(k+1)=\lambda_{k+1}(k)-2^kf(k+1), \nonumber \\
\lambda_{k+2}(k+1)=\lambda_{k+1}(k)+2^kf(k+1).
\end{eqnarray}
It is thus straightforward to check that the positivity and
(decreasing) monotonicity conditions mentioned above yield only non-negative
eigenvalues. We now turn to some concrete examples:

\begin{flushleft}
(i) The case of short-ranged spatial correlations.
\end{flushleft}

We consider first the function
\begin{eqnarray}
f(y)\ =\ {g \over (a_0+y)^\lambda }, 
\label{fshort}
\end{eqnarray}
with $\lambda>0$. In this case eq. (\ref{xqext}) becomes
\begin{eqnarray}
x(q)\ =\ {T\sqrt{2\ln K/g} \over
\{a_0^{-\lambda}-[a_0+L(1-q)]^{-\lambda}\}^{1/2}}
\label{xqshort}
\end{eqnarray}
and in the limit of large $L$ the solution for $x(q)$, $0<q<1$ becomes
\begin{eqnarray}
x(q)\ = \left\{ \begin{array}{ll}
{T \over T_c} & T \leq T_c \\
1 & T \geq T_c
\end{array} \right.
\label{xqs}
\end{eqnarray}
together with $x(0)=0$ and $x(1)=1$. Here $T_c$ is given by
\begin{eqnarray}
T_c\ =\ \sqrt{{f(0) \over 2\ln K}}.
\label{Tc}
\end{eqnarray}
For $T \leq T_c$, deviations of $x(q)$ from the form given in
expression (\ref{xqs}) are of
$O(L^{-\lambda})$, except when $1-q \sim O(1/L)$, and $x(q_c)$ becomes
equal to 1 for $q_c=1-O(1/L)$.

The solution found above is the same as the solution found in
ref. \cite{DS}, for zero-ranged correlations. Thus we see that the case
of short-ranged correlations is characterized by the same overlap
function and free energy as the zero-ranged case. The overlap
distribution function is given in terms of the solution for $x(q)$ by
\cite{MPV}
\begin{eqnarray}
P(q)= {dx(q) \over dq}.
\label{Pqx}
\end{eqnarray}
and hence
\begin{eqnarray}
P(q)\ = \left\{ \begin{array}{ll}{T \over T_c} \delta(q)\ + \ (\ 1\ -\
{T \over T_c}\ )\ \delta(q-1) & T \leq T_c \\ \delta(q) & T \geq T_c
\end{array} \right.,
\label{Pqs}
\end{eqnarray}
as discussed in the introduction. The free energy is obtained by
substituting the expression for $x(q)$ in eq. (\ref{bFfin}) and we
obtain:
\begin{eqnarray}
-F/L\ =\ \left\{ \begin{array}{ll}T_c \ln K + f(0) /(2 T_c) =\sqrt{2
f(0) \ln K} & T \leq T_c \\
T \ln K + f(0) /(2 T) & T \geq T_c \end{array} \right. ,
\label{Fsh}
\end{eqnarray}
where we have dropped corrections on the r.h.s which vanish as $L
\rightarrow \infty$. Note that on the Bethe lattice with the
hierarchical distance we did not find any substantial change as
$\lambda$ crosses the value 1 as has been found in the continuum limit.

\begin{flushleft}
(ii) The case of long-ranged correlations
\end{flushleft}

In this case we choose for the function governing the disorder
correlation
\begin{eqnarray}
f(y)\ =\ a_0\ -\ g \ y^\alpha,
\label{flr}
\end{eqnarray}
with $\alpha>0$ and $g>0$. 
We take $a_0$ to be very large, so that it satisfies
\begin{eqnarray}
T<<\sqrt{{a_0 \over 2 \ln K}}
\end{eqnarray}
for the range of temperatures we are interested in. Furtheremore, if
$L$ is such that it satisfies
\begin{eqnarray}
L<\left(a_0 \over g\right)^{1/\alpha},
\label{sml}
\end{eqnarray}
then $f(y)$ is positive for the entire range of allowed distances and
since it is a monotonically decreasing function of $y$ it is a
bona-fide correlation. The region of physical interest is $a_0
\rightarrow \infty$ with $L$ large but still satisfying the condition
(\ref{sml}). Eventually though, if one is interested in the true
thermodynamic limit $L \rightarrow \infty$ one has to consider the
case $L>(a_0/g)^{1/\alpha}$. In that case we must define
\begin{eqnarray}
f(y)= \left\{ \begin{array}{ll} a_0-g y^\alpha & \ \ 0 \leq y \leq
(a_0/g)^{1/\alpha} \\ 0 & \ \ (a_0/g)^{1/\alpha} \leq y \leq L
\end{array} \right.
\label{flrm}
\end{eqnarray}
for $f(y)$ to be a proper correlation function. We will comment about
the thermodynamic limit later in the section.

For now, starting with $f(y)$ given by eq. (\ref{flr}) with the condition
(\ref{sml}) being satisfied, the solution for $x(q)$ (see eq.
(\ref{xqext})) becomes
\begin{eqnarray}
x(q)\ = \left\{ \begin{array}{ll}
{T \sqrt{2\ln K} \over \sqrt{g L^\alpha (1-q)^\alpha}} & 0 \leq q \leq q_c \\
1 & q_c \leq q \leq 1 ,
\end{array} \right.
\label{xqlr}
\end{eqnarray}
with
\begin{eqnarray}
q_c\ =\ 1\ -\ \frac{1}{L} \left({T^2 2\ln K \over g} \right)^{1/\alpha}.
\label{qc}
\end{eqnarray}
We can identify an
$L$-dependent crossover temperature (at which $q_c=0$) given by
\begin{eqnarray}
T_c(L)\ =\ \left({g \over 2 \ln K}\right)^{1/2} L^{\alpha /2}
\label{TcL}
\end{eqnarray}
For $T$ fixed, as $L$ becomes large, $T_c(L)$ grows with $L$ and thus
we find ourselves always in the low temperature phase of the model,
characterized by $q_c>0$. (If on the other hand $T>T_c(L)$ then
$x(q)=1$ for any $q>0$). Assuming that $L$ is large enough so
$T<<T_c(L)$ (but the condition (\ref{sml}) still satisfied), we find by
substituting the expression for $x(q)$ in the formula for the free
energy, eq. (\ref{bFfin})
\begin{eqnarray}
-{\beta F \over L}\ =\ {T_c(L) \ln K \over T} \int_0^1 \ dq\
(1-q)^{\alpha/2} + {T \over T_c(L)} {gL^\alpha \over 2T^2} \int_0^1 \ dq\
(1-q)^{\alpha/2} +O(L^{-1})
\label{bFlr}
\end{eqnarray}
which can be simplified to give
\begin{eqnarray}
-F \ =\ {\sqrt{2g \ln K} \over 1+\alpha/2} L^{1+\alpha/2},
\label{Flr}
\end{eqnarray}
where we dropped constant terms. We see that the free energy is not
extensive for this range of $L$ values, but rather proportional to
$L^{1+\alpha/2}$. It is also 
temperature independent as it is in the low temperature phase of the
short-ranged case.

Let us now consider the distribution of overlaps. Using eq.
(\ref{Pqx}) we find
\begin{eqnarray}
P(q)\ = \left\{ \begin{array}{ll}{T \over T_c(L)} \delta(q)\ + \ 
{\alpha \over 2} {T \over T_c(L)} {1 \over (1-q)^{1+\alpha/2}}
& 0 \leq q \leq q_c \\ 0 & q_c \leq q \leq 1
\end{array} \right.,
\label{Pqlr}
\end{eqnarray}
in the limit of large $L$ this expression becomes simply
\begin{eqnarray}
P(q) = \delta(1-q).
\label{Pql}
\end{eqnarray}
This becomes obvious by going back to the expression for $x(q)$ which
in the limit of large $L$ become $x(q)=0$ for $0 \leq q <1$ and
$x(1)=1$, which can also be expressed as 
\begin{eqnarray}
q(x)= \left\{ \begin{array}{ll} 0 & x=0 \\ 1 & 0 <x \leq 1 \end{array}
\right. 
\label{qx}
\end{eqnarray}
Thus in the limit of large $L$ the solution becomes replica symmetric for
any fixed temperature.

Before we close this section we show that in the true thermodynamic
limit the free energy becomes extensive. To achieve this we must 
allow $L>(a_0/g)^{1/\alpha}$ and use the correlation function defined
by eq.(\ref{flrm}). We still demand that $a_0$ be very large so
$T<<\sqrt{a_0/ (2 \ln K)}$ is always satisfied. In that case we
find for $x(q)$:   
\begin{eqnarray}
x(q)=\left\{\begin{array}{ll}T/T_c(a_0) & \ \ 0 \leq q \leq q_{c1} \\
{T \sqrt{2 \ln K} \over \sqrt{gL^\alpha(1-q)^\alpha}} & \ \ q_{c1}
\leq q \leq q_{c2} \\ 1 & \ \ q_{c2} \leq q \leq 1 \end{array} \right.
\end{eqnarray}
where
\begin{eqnarray}
T_c(a_0)=\left(a_0 \over 2\ln K \right)^{1/2}, \\
q_{c1}=1-{1 \over L}\left(a_0 \over g\right)^{1/\alpha}, \\
q_{c2}=1-{1 \over L}\left(T^2 2\ln K \over g\right)^{1/\alpha}.
\end{eqnarray}
In the limit $L \rightarrow \infty$ one finds
\begin{eqnarray}
x(q)=T/T_c(a_0), \ \ 0 <q<1,
\end{eqnarray}
together with $x(0)=0$ and $x(1)=1$. In the limit $T<<T_c(a_0)$ we see
that practically $x(q)=0$ for $0\leq q<1$ and $x(1)=1$ which amounts
to $P(q)=\delta(1-q)$ as before. The free energy though is given by
\begin{eqnarray}
-F/L=\sqrt{2a_0\ln K}
\end{eqnarray}
and it is thus an extensive function of L as it should be in the
thermodynamic limit.
 
\section{ {\bf Summary and discussion }}
\label{sec:5}
 
In this paper we have considered the case of directed polymers on a
Cayley tree in the presence of correlated disorder. We have used the
ultrametric hierarchical distance to introduce distance within each
shell, and this distance is simple enough to enable us to solve the
model exactly under the assumption of a hierarchical Parisi-type
solution.

We have found two different types of behavior depending on the range
of the disorder correlations. In the case of short range correlations
the solution behaves like the non-correlated case: There is a phase
transition at a finite temperature and the two phases differ by the
temperature dependence of the free energy and by the expression for
the overlap distribution which is non trivial in the low temperature
phase.  In the case of long range correlations there is no phase
transition as a function of temperature (strictly speaking the
transition temperature $T_c(a_0)$ can be made as large as we please by
choosing $a_0$ to be large enough). This is similar to the behavior
that has been found in the continuum limit \cite{mp}. However we have
identified an $L$-dependent crossover temperature which plays a role
for a finite-size system. We have also found
that in the large $L$ limit the solution becomes replica-symmetric but
with the overlap distribution peaked at $q=1$ at any temperature,
which is the case in the short-ranged case only at $T=0$. This result
is reminiscent of Parisi's solution \cite{parisi2} for the case of
harmonic correlations in the continuum limit where no RSB has been
found. This is in contrast to results in the continuum limit for the
non-harmonic case \cite{mp} where the variational approximation
yielded an infinite-step RSB for the long-ranged case (but also no
phase transition). Another feature we have found in the long-ranged
case, which is similar to Parisi's result \cite{parisi2}, is the
non-extensivity of the free energy in terms of $L$ over a large range
of $L$-values. Eventually as $L \rightarrow \infty$, the free energy
become extensive. The
exponent $1+\alpha/2$ may be related to the
exponent $\omega$ which characterizes the free energy fluctuations, see
eq. (\ref{omega}), but this can be established only after carrying
out a calculation of the free energy fluctuations on the Cayley tree.

We should emphasize, that because of the non-euclidean nature of the
hierarchical distance on a tree, we could not establish a relation
between the exponents $\lambda$ in eq. (\ref{fshort}) or $\alpha$ in
eq. (\ref{flr}) to the exponent $\gamma$ defined in eqs. (\ref{Fys})
and (\ref{Fyl}) which characterizes the range of disorder correlations
on ordinary 
lattices embedded in euclidean space. Related to this is the fact that
the separation between short and long range correlations occurs for
$\gamma=2$ for ordinary lattices (see eq.(\ref{Fys}) and ref. \cite{mp})
whereas we find the separation to occur at $\lambda =0$ 
(or $\alpha=0$) for the tree problem. 

There are various possibilities to extend this work further. One is to
consider the transverse distance defined in section \ref{sec:2}, and
attempt to solve the model, including the behavior of the
root-mean-square transverse distance characterized by an exponent
$\nu$.

Also, we have only considered the case of a gaussian distribution of
the disorder. Other distributions are possible, some better suited for
calculating $1/d$ corrections (like the exponential distribution)
\cite{c&d2}. One can attempt to obtain these corrections for the case
of long-ranged correlations of the disorder.

{\bf Acknowledgements:}
I thank T. Blum for a useful discussion and some comments. I thank the
referee for some useful comments and suggestions. This work is partially
supported by the US Department of Energy, grant No. DE-G02-98ER45686.

\begin{figure}
  \centerline{\epsfysize 5.5cm \epsfbox{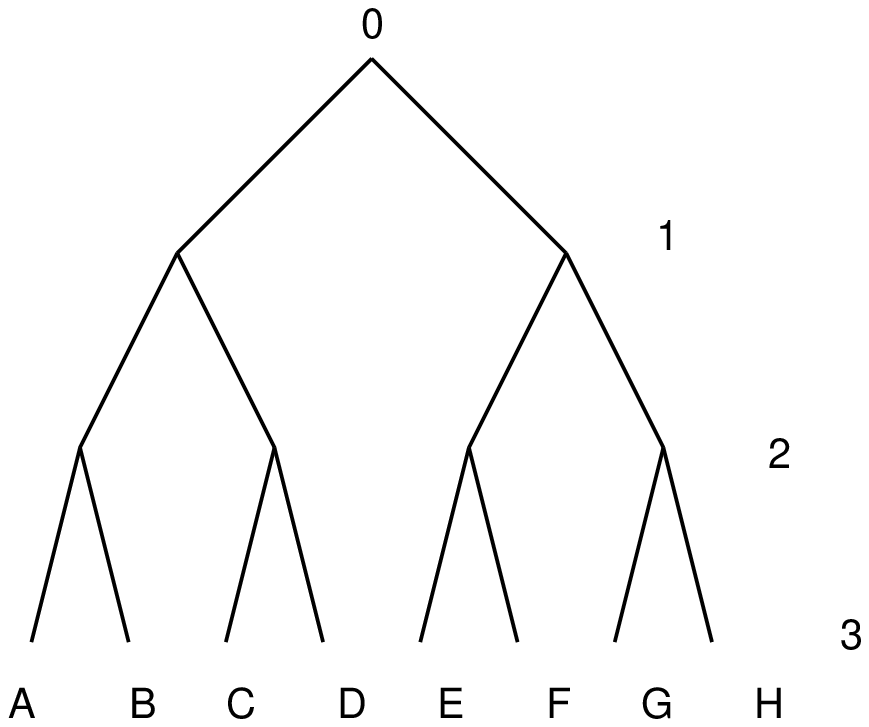}} \vskip 0.3 truecm
\caption{A branch of a tree with branching ratio K=2. The
ultrametric distance between points A and B is 1, between A and C (or D)
is 2 and between A and E (or A and  F, G, H) is 3.}
\label{fig1}
\end{figure}

\end{document}